\documentclass[12pt]{iopart}

\usepackage{graphicx}
\usepackage[dvipsnames]{xcolor}
\newcommand{\Dafne }{DA$\Phi$NE}
\PassOptionsToPackage{hyphens}{url}\usepackage{hyperref}
\usepackage{lineno}

\begin{document}

\title{Kaonic Atoms at the \Dafne{} Collider with the SIDDHARTA-2 Experiment }
\author{ 
F Napolitano$^{1}$,
F Sgaramella$^{1}$,
M Bazzi$^{1}$, 
D Bosnar$^{2}$,
M Bragadireanu$^{3}$, 
M Carminati$^{4}$,
M Cargnelli$^{5}$, 
A Clozza$^{1}$, 
G Deda$^{4}$,
L De~Paolis$^{1}$, 
R Del Grande$^{1,6}$,
L Fabbietti$^{6}$,
C Fiorini$^{4}$,
C Guaraldo$^{1}$,
M Iliescu$^{1}$,
M Iwasaki$^{7}$,
P Levi Sandri$^{1}$
J Marton$^{5}$,
M Miliucci$^{1}$,
P Moskal$^{8}$,
S Nied\`zwiecki$^{8}$
K Piscicchia$^{9}$,
A Scordo$^{1}$,
H Shi$^{5}$, 
D Sirghi$^{1}$,
F Sirghi$^{1}$,
M Silarski$^{8}$,
M Skurzok$^{8}$,
A Spallone$^{1}$,
M Tüchler$^{5}$,
J Zmeskal$^{5}$ and
C Curceanu$^{1}$
}

\address{$^{1}$ INFN, Laboratori Nazionali di Frascati, {Via E. Fermi 54, I-00044 Roma, Italy}}
\address{$^{2}$ Department of Physics, Faculty of Science, University of Zagreb, {10000 Zagreb, Croatia}}
\address{$^{3}$ Horia Hulubei National Institute of Physics and Nuclear Engineering, {IFIN-HH, 077125 Magurele, Romania}}
\address{$^{4}$ Politecnico di Milano, Dipartimento di Elettronica, Informazione e Bioingegneria and INFN Sezione di Milano, {20133 Milano, Italy}}

\address{$^{5}$ Stefan-Meyer-Institute for Subatomic Physics, Austrian Academy of Science, {Kegelgasse 27, 1030 Vienna, Austria }}
\address{$^{6}$ Excellence Cluster Universe, Technische Universit\"{a}t M\"{u}nchen, { Boltzmannstra{\ss}e 2, D-85748 Garching, Germany}}
\address{$^{7}$ RIKEN {Tokyo 351-0198, Japan}}
\address{$^{8}$ The M. Smoluchowski Institute of Physics, Jagiellonian University, 30-348 Kraków, Poland }
\address{$^{9}$ Centro Ricerche Enrico Fermi--Museo Storico della Fisica e Centro Studi e Ricerche ``Enrico Fermi'', {Piazza del Viminale 1, I-00184 Roma, Italy} }
\ead{Napolitano.Fabrizio@lnf.infn.it, Francesco.Sgaramella@lnf.infn.it}


\begin{abstract}
Kaonic atoms are a unique tool to explore quantum chromodynamics in the strangeness sector at low energy, with implications reaching neutron stars and dark matter. Precision X-ray spectroscopy can fully unlock the at-threshold isospin dependent antikaon-nucleon scattering lengths, via the atomic transitions to the fundamental level. While the SIDDHARTA experiment at the INFN-LNF \Dafne{}  collider successfully measured kaonic hydrogen, its successor SIDDHARTA-2 is starting now its data taking campaign aiming to finally fully disentangle the isoscalar and isovector scattering lengths via the measurement of kaonic deuterium. An overview of the first experimental results from a preparatory run for the SIDDAHARTA-2 experiment is presented.
\end{abstract}
\submitto{\PS}

\section{Introduction}
Exotic bound states of atomic systems have been studied for more than 80 years~\cite{PhysRev.58.90.2}; they can be formed by sufficiently long-lived negatively charged particles. This is the case for both leptonic, such as muons, or hadronic particles as pions, kaons, anti-protons or sigma hyperons. The study of these systems paved the way to a deeper understanding of particle and nuclear physics and aided the development of the Standard Model. Hadronic exotic atoms in particular can reach low-energy regimes of the strong interaction, which can hardly be achieved by scattering experiments. The hadron-nucleus interaction in exotic atoms occurs at negligible momentum transfer, nearly `at-threshold'. The quantum chromodynamics (QCD) in the strangeness sector can be probed directly at low energy in kaonic atoms.

QCD describes the strong interaction of quarks and gluons. Contrarily to Quantum Electrodynamics, which exhibits a screening of the charge due to vacuum polarization, gluons of QCD vacuum are self-interacting, leading to anti-screening. At higher momentum transfer (or equivalently at smaller distances), the strength of the interaction is smaller; quarks and gluons behave almost as free particles, enabling the treatment by perturbation theory (asymptotic freedom). At lower momentum transfer (or larger distances) on the other hand, the strength is larger. This leads to a peculiar characteristic known as confinement: quarks are confined inside the nucleons due to the stronger force they experience at small momenta. Below energies of  $\sim$1 GeV, perturbative QCD cannot be employed anymore and simplified, phenomenological models are used instead. 
The understanding of $\bar{K}N$ interactions in this regime is based on Chiral Perturbation Theory ($\chi$PT); however, simple approaches fail due to the strong coupling to the $\pi \Sigma$ channel and the presence of the $\Lambda$(1405) resonance just below threshold~\cite{1994}~\cite{2008}. A comparative analysis of the state-of-the-art theoretical approaches is presented in~\cite{ciepl2016pole}. The majority of these models (Bonn, Murcia and Kyoto-Munich) exploit dimensional regularization to keep under control the ultraviolet divergences; Prague model uses off-shell form factors. 
All the models employ free parameters fitted to the low energy $\bar{K}p$ experimental data which are available through the study of kaonic hydrogen. However, great discrepancy remains in the predictions of the $\bar{K}n$ scattering lengths; this can only be accessed via a measurement of kaonic deuterium, which is the main goal of the SIDDHARTA-2 experiment, as discussed below.

\subsection{Kaonic Atoms}
\begin{figure}[h]
\centering
\includegraphics[width=0.8\textwidth]{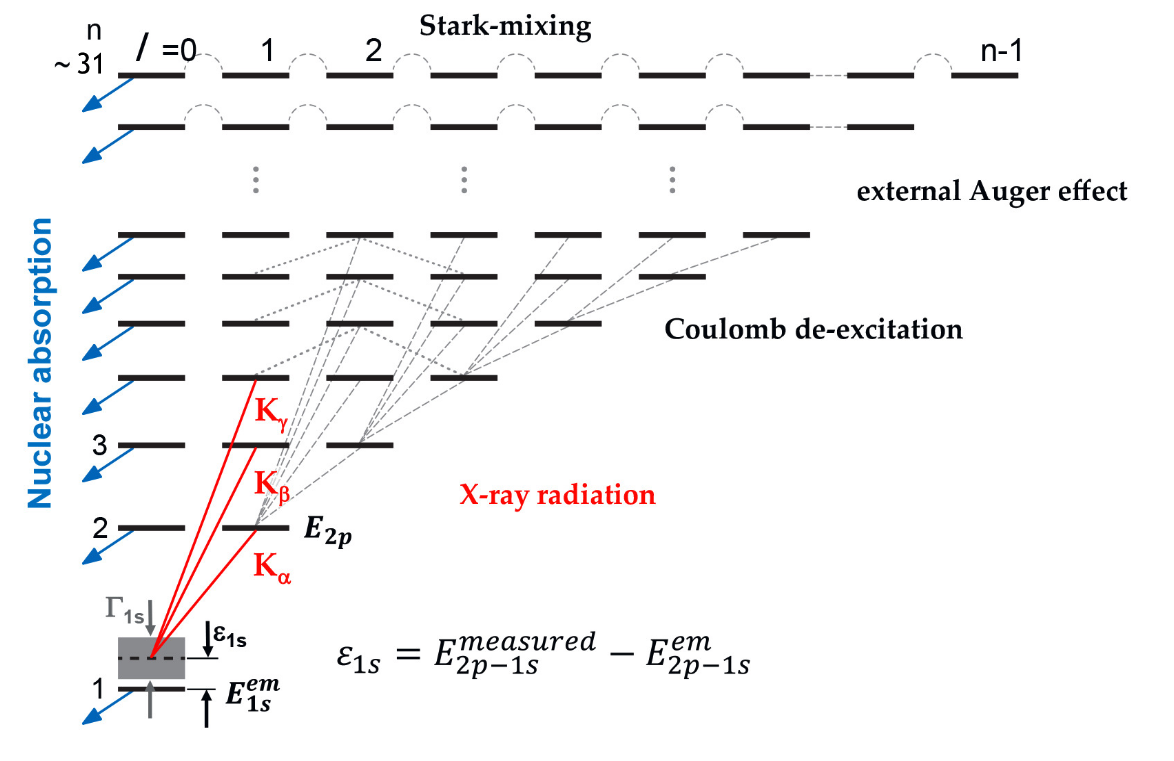}
\caption{A schematic overview of the de-excitations which take place during the formation of the kaonic atoms. The kaons are captured in a highly excited state and cascade down to the fundamental level. The last transitions are influenced by the strong interaction, leading to a broadening and shift of the 1s level with respect to the case of only electromagnetic interaction~\cite{10.21468/SciPostPhysProc.3.039}.}
\label{fig00}
\end{figure}

A kaonic atom is formed when a kaon is slowed down, or has an initial momentum low enough to be stopped in a target and be captured by an atom via the electromagnetic interaction. This occurs at a highly excited state~\cite{zmeskal2008kaonic}, which depends on the reduced mass of the system. In kaonic hydrogen this takes place at $n\approx$ 25. After capture, the system starts de-exciting towards the fundamental level, emitting radiation in the X-ray domain.
Due to the Stark effect and other competing processes which increase with density, the yield of kaons which can de-excite completely to the ground state strongly depends on the target material.
The strong interaction causes an energy shift of the transition to the fundamental 1s level with respect to the canonical electromagnetic one, $E^{em}_{2p\to 1s}$. An overview of these processes is outlined in Figure~\ref{fig00}.

The vicinity of the kaon to the nucleus is such that the strong interaction is not negligible anymore, and this impacts on the energy of the radiation emitted $E^{measured}_{2p\to 1s}$. This shift, $\epsilon_{1s}$, is defined as:
\begin{equation}
\epsilon_{1s} = E^{measured}_{2p\to 1s} - E^{em}_{2p\to 1s} 
\end{equation}
Together with the energy shift, strong interaction causes also a broadening of the X-ray energy of the emission 2p$\to$1s, measured via the width, $\Gamma_{1s}$.

The $\bar{K}p$ scattering length, $a_{\bar{K}p}$, is connected to $\epsilon_{1s}^{H}$ and $\Gamma_{1s}^{H}$ via the Deser-Trueman formula~\cite{trueman1961energy}. In the improved version by Meissner et al.~\cite{meissner2006kaon}, which takes into account the isospin-breaking corrections, this is written as:

\begin{equation}
 \epsilon_{1s}^{H}+\frac{i}{2}\Gamma_{1s}^{H} = 2 \alpha^3 \mu^2 a_{\bar{K}p} \left( 1-2\alpha\mu(\texttt{ln} \alpha -1 ) a_{\bar{K}p} + ... \right) 
\end{equation}
Similarly, the $\bar{K}d$ scattering length, $a_{\bar{K}d}$, is linked to the kaonic deuterium shift and width:

\begin{equation}
\epsilon_{1s}^{D}+\frac{i}{2}\Gamma_{1s}^{D} = 2 \alpha^3 \mu^2 a_{\bar{K}d} \left( 1-2\alpha\mu(\texttt{ln} \alpha -1 ) a_{\bar{K}d} + ... \right)
\end{equation}

where $\alpha$ is the fine structure constant and $\mu$ the reduced mass of the system.

The $\bar{K}p$ scattering length is connected to the $\bar{K}N$ isospin-dependent ($I=1,0$) scattering lengths $a_0$ and $a_1$ via the relation:
\begin{equation} \label{eq:1}
\centering a_{\bar{K}p} = \frac{1}{2} (a_0 + a_1)
\end{equation}

The individual isoscalar $a_0$ and isovector $a_1$ scattering lengths can be obtained via the measurement of kaonic deuterium. This in fact provides information on a different combination of $a_0$ and $a_1$:

\begin{eqnarray}
a_{\bar{K}n} =& a_1\label{eq:2}\\
a_{\bar{K}d} =&\frac{4(m_N+m_K)}{2 m_N+m_K}Q + C \label{eq:3}
\end{eqnarray}

Where $Q = 1/2(a_{\bar{K}p}+a_{\bar{K}n}) = 1/4 (a_0+3a_1) $. The first addend in eq.~\ref{eq:3} represents the lowest order impulse approximation. 
The second term $C$ includes all higher order contributions, namely all other physics associated with the $\bar{K}d$ three-body interaction.
The three-body system $\bar{K}NN$ can be studied by solving Faddeev-type equations. Since the $\bar{K}d$ three-body problem includes the complication that the $\bar{K}p$ and $\bar{K}n$ interactions involve significant inelastic channels, one has to use a coupled-channel formalism, including both elastic and inelastic channels.

The experimental effort to determine shift and width of kaonic deuterium is challenged by its lower yield with respect to kaonic hydrogen. This challenge requires an up-scaling of the detector technologies, as detailed below.

\section{Kaonic atoms experiments}
The experimental effort for the study of the kaonic exotic atoms takes place at the INFN-LNF research centre, and at J-PARC. The two complexes employ very different experimental techniques and strategies,  based on the different characteristics of the accelerators.
The \Dafne{} complex at INFN-LNF~\cite{Milardi:IPAC2018-MOPMF088}~\cite{vignola} is an $e^+e^-$ collider, and started operations in 1997. The centre of mass energy is set at the $\phi$ resonance, 1.02 GeV, so to provide kaons through the process $e^+e^-\to \phi \to K^+K^-$ which is the decay channel with the highest branching ratio of the $\phi$ meson (48.9$\pm$0.5\%)~\cite{10.1093/ptep/ptaa104}. The collider, after the crab-waist scheme upgrade~\cite{PhysRevLett.104.174801}, reached a peak luminosity of 4.5$\times$10$^{32}$ cm$^{-2}$s$^{-1}$.
The charged kaons from the decays have a small, almost mono-energetic momentum of 127 MeV/c ($\Delta p / p = 0.1\%$), which means they can stop efficiently inside a gaseous target with small material budget. 
\begin{figure}[h]
\centering
\includegraphics[width=0.8\textwidth]{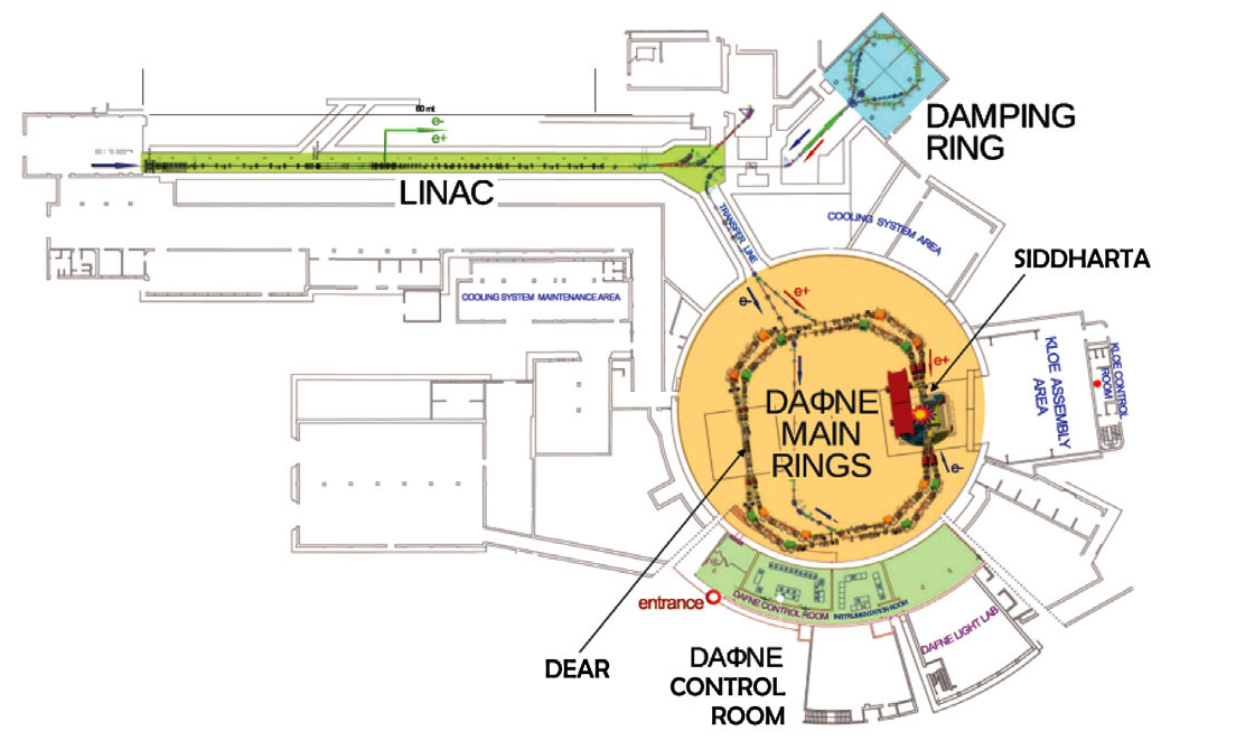}
\caption{Outline of the \Dafne{} accelerator complex with its main rings, the damping ring and the linear accelerator, LINAC (\cite{RevModPhys.91.025006} and references therein). }
\label{fig0}
\end{figure}

The overview of the \Dafne{} collider, the LINAC and the damping ring are shown in Figure~\ref{fig0}. 
The production of kaonic atoms at an $e^+e^-$ collider has the advantage of a lower hadronic background; however, due to kaons being produced from the $\phi$ decay, the acceptance of the target is also a critical parameter. The SIDDHARTA experiment operated at \Dafne{}, solving a decade long puzzle of incompatible measurements of kaonic hydrogen~\cite{BAZZI2011113}, and producing the first observation of kaonic helium-3~\cite{SIDDHARTA:2010uae} and measuring kaonic helium-4~\cite{SIDDHARTA:2009qht}. Its successor, the SIDDHARTA-2 experiment, aims at the first measurement of kaonic deuterium, as it is detailed below. 

By contrast, the KEK Japan Proton Accelerator Research Complex (J-PARC)~\cite{2010} makes use of a proton beam with high intensity, serving multiple experiments and purposes through the delivery of secondary beams. From the primary 30 GeV protons, a kaon beam is extracted at an energy of about 1 GeV. Major present experiments are the E62, which is dedicated to the measurement of kaonic helium and the E57 experiment for kaonic deuterium~\cite{2020jparc}. 
They take advantage of the high-intensity secondary kaon beam. The higher energy and the harder hadronic radiation environment with respect to the conditions present at the \Dafne{} collider, however, represent major experimental challenges. 

\section{The SIDDHARTA-2 Experiment}
The SIDDHARTA-2 (SIlicon Drift Detector for Hadronic Atom Research by Timing Application) experiment is currently installed at the INFN-LNF \Dafne{} collider, and is starting its data taking campaign for the measurement of kaonic deuterium. During the commissioning phase with a pilot run with a reduced setup, SIDDHARTINO, the experimental conditions and background sources present at the collider were assessed. The run successfully concluded in summer 2021 with a kaonic helium data taking. 

\begin{figure}[h]
\centering
\includegraphics[width=0.8\textwidth]{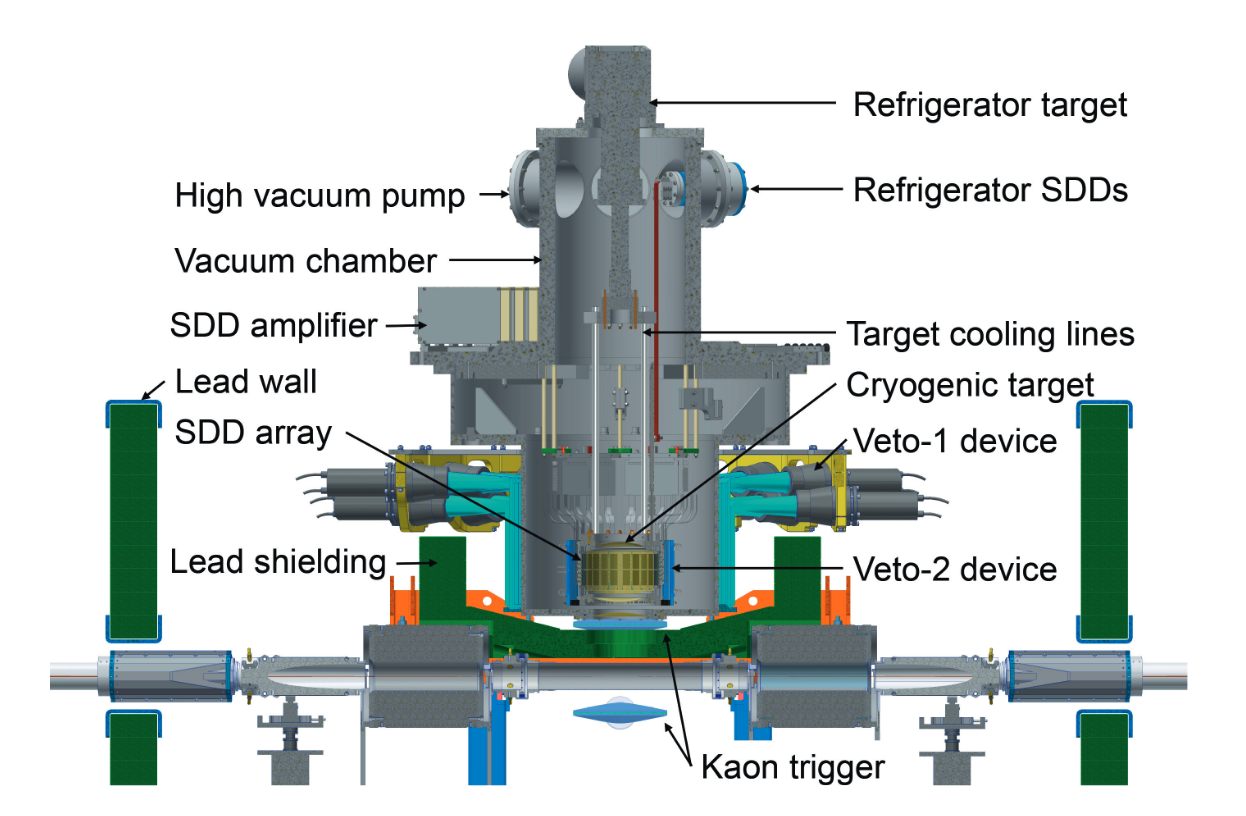}
\caption{Schematics of the SIDDHARTA-2 experiment at \Dafne{}. Details of the apparatus in the text.}
\label{fig1}
\end{figure}

\begin{figure}[h]
\centering
\includegraphics[width=0.8\textwidth]{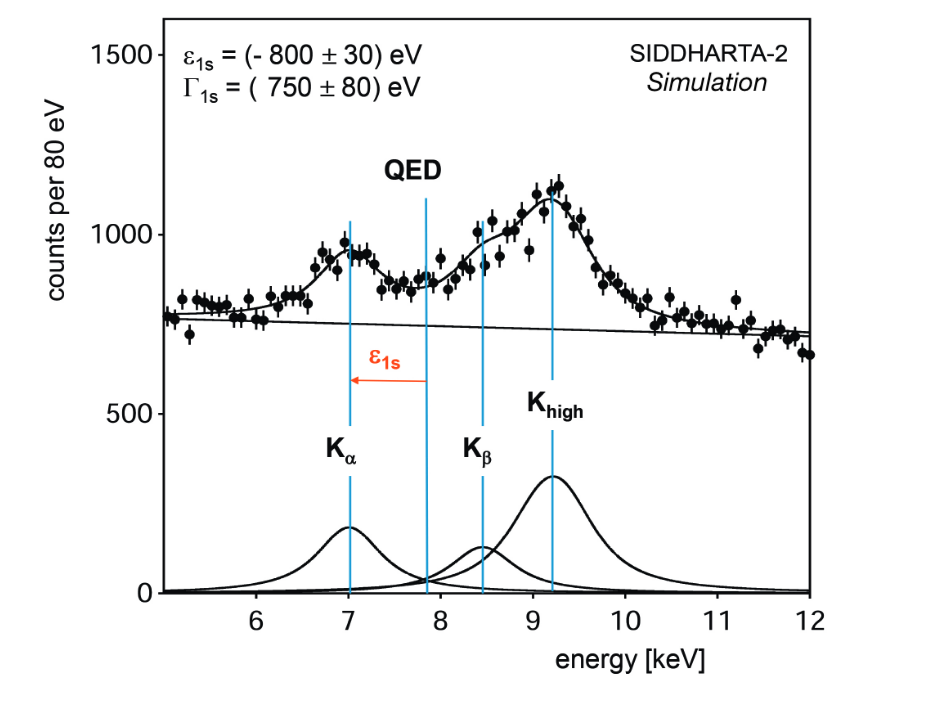}
\caption{Monte-Carlo simulation of the expected SDD spectrum for the kaonic deuterium measurement, corresponding to 800 pb$^{-1}$ of integrated luminosity~\cite{10.21468/SciPostPhysProc.3.039}. The data was generated assuming 800 eV of shift and 750 eV of width, with a K$_\alpha$ yield of 10$^{-3}$. }
\label{fig12}
\end{figure}

In Figure~\ref{fig1}, the details of the experimental apparatus are shown. Above and below the Interaction Region (IR), two pairs of plastic scintillators read by photo-multipliers act as Kaon Trigger, by means of time-of-flight. The purpose of the Kaon Trigger is to select kaons which are emitted almost back-to-back from the decay of the $\phi$ meson in the IR and directed towards the target.
A cylindrical vacuum chamber is placed above the Interaction Point (IP) of the IR and contains the target cell.
The kaons continue inside the vacuum chamber, where they interact with the atoms of the gas of the target to form kaonic atoms and emit X-rays. 
Twelve pairs of plastic scintillators read by photo-multiplier tubes, the Veto-1 system~\cite{Bazzi_2013}, are placed around the cryogenic target, outside the vacuum chamber. Inside the vacuum chamber, radially around the target, smaller scintillators read by silicon photo-multipliers (SiPMs) are used as an additional veto system (Veto-2)~\cite{T_chler_2018}. The Veto-1 and Veto-2 systems are used to suppress the synchronous and asynchronous background from the accelerator, and limit the fake signals due to minimum ionizing particles (MIPs).
Downstream and upstream the beam-pipe, lead shielding is placed around the vacuum chamber to minimize the background from the accelerator. 
The cryogenic target is constructed with Kapton walls, with an aluminium structure. Around the target, the Silicon Drift Detectors (SDDs) are used to detect the X-rays produced in the decays of the kaonic atoms.
The SDDs and their front-end electronics are described in more details in~\ref{sub:silicon_drift_detectors}.
The setup is equipped with two X-ray tubes which are used to determine the energy calibration.
The high vacuum pump and the cryogenic system connectors are placed on top of the vacuum chamber. Between the target and the IR, an additional material budget is placed, the degrader, to optimize the stopping range of the kaons within the gaseous target.
To provide luminosity feedback to \Dafne{}, another pair of plastic scintillators read by photo-multiplier tubes is placed on the longitudinal plane, in front of the IR. The SIDDHARTA-2 Luminometer~\cite{Skurzok_2020} uses the kaon rates to measure the luminosity. This monitor is complementary to the \Dafne{} luminosity monitor, which, instead, uses Bhabha scattering $e^{+}e^{-}\rightarrow e^{+}e^{-}$. 
The geometry of the entire apparatus was optimized with a \texttt{GEANT4}~\cite{AGOSTINELLI2003250} simulation, such as to assure the highest signal yield. This implies in particular a study of the degrader thickness. 
Taking into account the much lower yield of  kaonic deuterium with respect to kaonic hydrogen (about 1/10), 800 pb$^{-1}$ of integrated luminosity are needed to achieve a precision comparable to the SIDDHARTA measurement of kaonic hydrogen. The simulation of kaonic deuterium signal is shown in Figure~\ref{fig12}. The K$_\alpha$, K$_\beta$ and the higher order transitions (K$_{high}$) are indicated on the plot, together with the position of the purely electromagnetic energy transition.

\subsection{Silicon Drift Detectors}
\label{sub:silicon_drift_detectors}
State-of-the-art resolution and radiation hardness is a key requirement in precision X-ray spectroscopy, which is needed for the measurement of the kaonic deuterium. Another key requirement is a large number and surface area of the SDDs, in order to maximize the signal yield.

Custom monolithic SDD detectors have been developed by a collaboration of Politecnico di Milano, Fondazione Bruno Kessler, Stefan Meyer Institute and INFN-LNF~\cite{condmat6040047}~\cite{QUAGLIA2016449}.  To satisfy the requirements, the SDDs have a thickness of 450 $\mathrm{\mu}$m and are organized in arrays of 2 $\times$ 4 matrices of 5.12 cm$^2$, with a 75\% active area. Their quantum efficiency is above 85\% in the energy range considered for the transitions. The SDDs are designed to operate at 170 K, close to the cryogenic target. A low-noise charge sensitive preamplifier (CUBE)~\cite{7466864} represents the first stage of the signal processing, together with the front-end readout based on the SFERA ASIC. These can provide charge and timing information to the downstream data acquisition system. The timing measurement in particular is essential to suppress asynchronous background which originates from the accelerator. The SDDs have been extensively tested during the SIDDHARTINO run in the hard radiation environment of \Dafne{}, showing optimal stability, linearity and resolution of both timing and energy. The energy resolution of the SDDs was determined to be around 158 eV at the Fe K$_\alpha$ line and the X-ray response linear within a few eV~\cite{Miliucci_2021}. From a total of 48 SDDs arrays, 8 have been used in the commissioning SIDDHARTINO run, while the remaining detectors were fully installed before the start of the kaonic deuterium data taking campaign.

\section{First Results From The Commissioning Phase}

During the SIDDHARTINO phase from January 2021 to July 2021, the \Dafne{} collider delivered an integrated luminosity of more than 50 pb$^{-1}$ which was used to assess the setup readiness and the background levels for the upcoming SIDDHARTA-2 runs. The target was filled with helium gas at different densities; the kaonic helium emission L$_\alpha$ line was measured in the X-ray spectrum~\cite{condmat6040047} and used to validate the optimal degrader thickness predicted by a  Monte-Carlo simulation. 
The use of the Kaon Trigger and the SDDs timing alone showed a background rejection factor of about $10^5$.

\begin{figure}[h]
\centering
\includegraphics[width=0.8\textwidth]{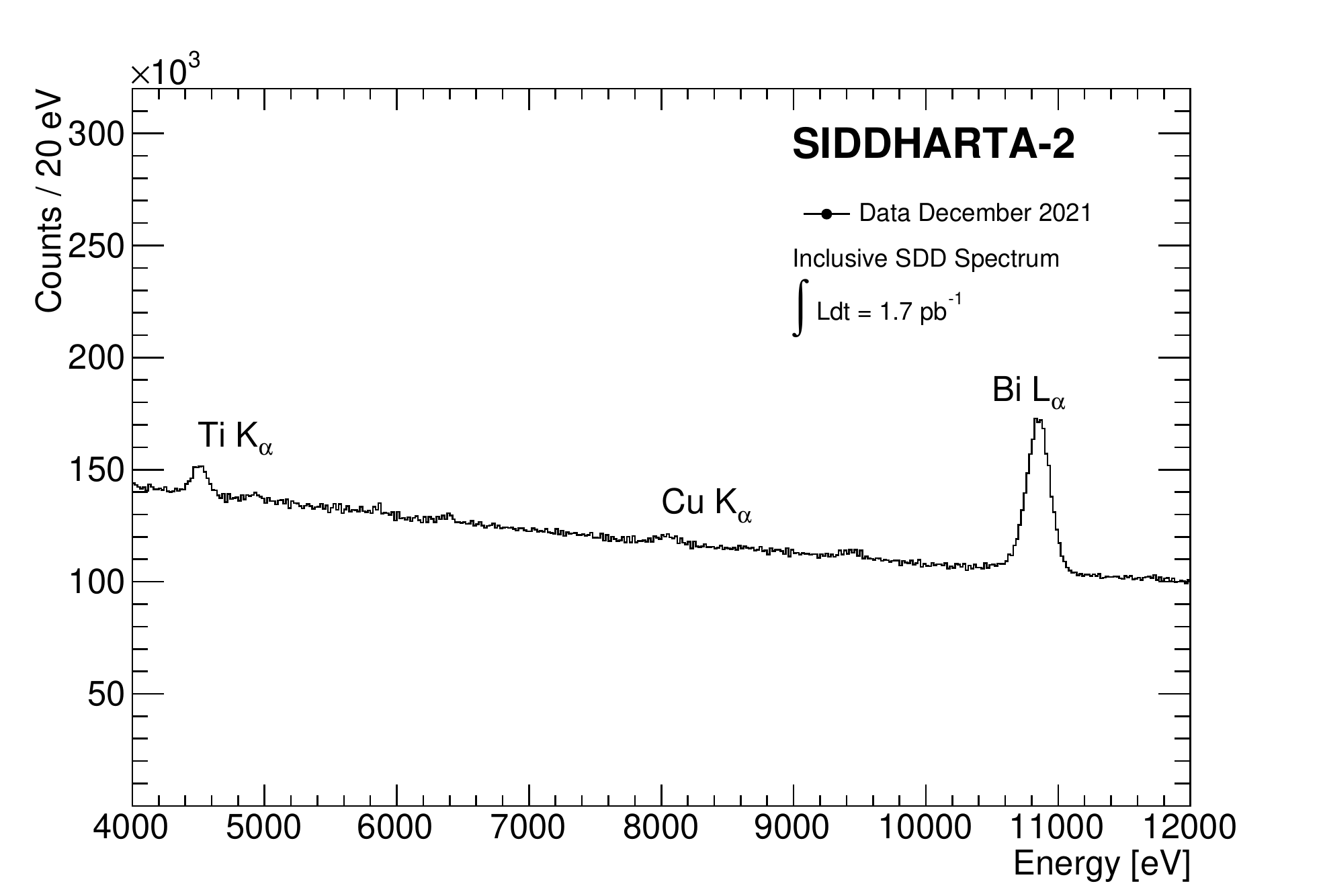}
\caption{SIDDHARTA-2 SDD energy inclusive spectrum using 1.7 pb$^{-1}$ of $e^+ e^-$ data delivered by DA$\Phi$NE. The inclusive spectrum shows the activation of the titanium, copper, and bismuth lines from the radiation emitted by the accelerator.}
\label{fig2m1}
\end{figure}

After summer 2021, the Veto-1 and Veto-2 were installed in the setup, together with additional SDD arrays and upgraded shielding. During this final commissioning phase of SIDDHARTA-2 before the kaonic deuterium measurement, the emission spectrum from the helium target (at 1.48\% liquid density) was acquired. The inclusive spectrum is shown in Figure~\ref{fig2m1}, which corresponds to the data collected by the SDDs prior to any requirement. The spectrum shows spectroscopic peaks which are related with the activation of material around the detectors. Titanium and Copper are placed inside the vacuum chamber close to the SDDs for calibration purposes (Ti K$_\alpha$ and Cu K$_\alpha$). The bismuth, present in the SDD detector ceramics, is activated as well (Bi L$_\alpha$).

\begin{center}
\begin{table}[h!]
\centering

\begin{tabular}{l c c}
 \hline
 Selection & Events in ROI & Rejection Factor \\ 
 \hline \hline
SDD events & 48739731 &  -\\
Injection cut & 28811141 & 5.9$\times$10$^{-1}$\\
Trigger cut & 1554 & 5.4 $\times$10$^{-5}$\\
T0F cut & 1110 & 9.6$\times$10$^{-1}$\\
Drift cut & 596 & 5.4$\times$10$^{-1}$\\
 \hline
Total & 596 &  1.2$\times$10$^{-5}$\\
 \hline
\end{tabular}

\caption{Outline of the selection requirements applied to the data and the corresponding number of events passing each cut in the Region Of Interest (ROI) from 4 to 12 keV, with the associated rejection factors. The cuts are described in the text.}
\label{table:1}

\end{table}
\end{center}

In Table~\ref{table:1}, the overview of the selections applied to the collected data, and the number of events passing each requirement together with the rejection factor are given.
The event selection makes use of the additional information provided by the SIDDHARTA-2 sub-detectors as well as the SDDs. 
Events taken during \Dafne{} injection are typically affected by much higher background conditions, and therefore discarded (Injection cut). 
If the two pairs of photo-multiplier tubes of the Kaon Trigger detect signals, then the event is marked as triggered (Trigger cut). This cut is extremely efficient to reject the background events, since the position of the scintillators in the transverse plane assures that the out-of-bunch (asynchronous) background coming from the collider rings is mostly suppressed. After this cut, only particles coming from the IR and their interaction with the apparatus contribute to the background. To select only the events in temporal coincidence with the kaons, the timing of the scintillator's signal is calculated to determine the time of flight of the particle, which has to be compatible with that of the kaons from the IR (TOF cut). Finally, the drift time of the SDDs is selected to be within the timing window of the  trigger (Drift cut).

\begin{figure}[h]
\centering
\includegraphics[width=0.8\textwidth]{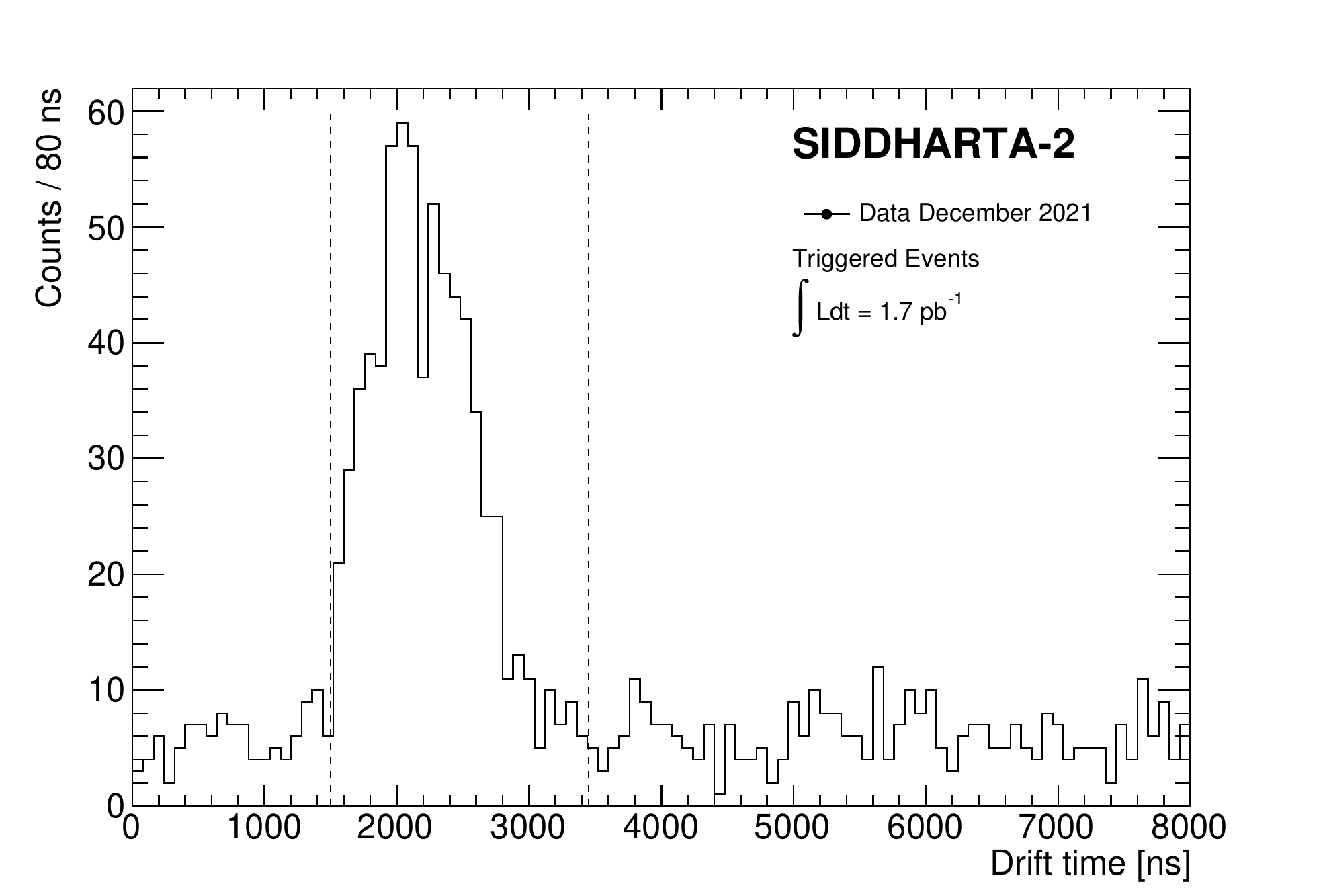}
\caption{SDD timing spectrum after the trigger selection. The peak at the centre of the spectrum are the events in close temporal proximity with the kaonic trigger signal. The dashed black lines represent the cut used to select the kaonic events (Drift cut).}
\label{fig21}
\end{figure}

The Drift cut is shown in Figure~\ref{fig21} for events passing the trigger requirement. This final cut further reduces the synchronous background, increasing the purity of the selected data.

\begin{figure}[h]
\centering
\includegraphics[width=0.8\textwidth]{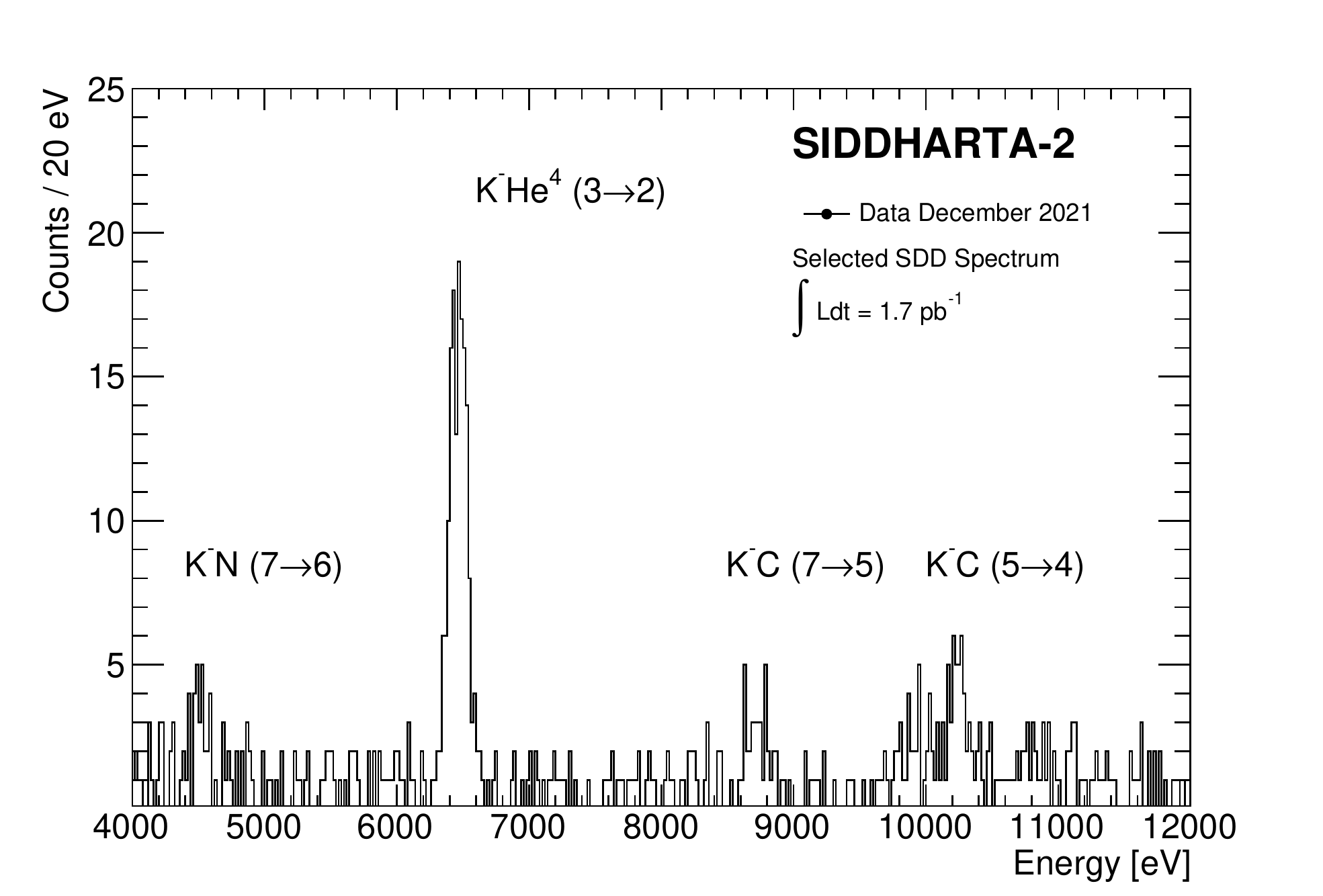}
\caption{SIDDHARTA-2 SDD energy spectrum after selection using 1.7 pb$^{-1}$ of $e^+ e^-$ data delivered by \Dafne{}. Transitions from kaonic atoms are visible, the kaonic helium peak at around 6.5 keV is the most prominent; kaonic carbon and kaonic nitrogen transitions are present as well. The cryogenic target was filled with helium at 1.48\% liquid density.}
\label{fig2}
\end{figure}

In Figure~\ref{fig2} the SDDs spectrum in the Region Of Interest for the kaonic helium (4 keV to 10 keV), corresponding to the commissioning run of the SIDDHARTA-2 setup in December 2021, with a total integrated luminosity of 1.7 pb$^{-1}$, is shown. In the spectrum it is visible the prominent kaonic helium L$_\alpha$ transition, together with transitions of kaonic nitrogen and kaonic carbon. These additional kaonic atoms are generated by the interaction of the kaons with the Kapton walls of the cryogenic target.

\section{Conclusions}
The measurement of shift and width of the kaonic deuterium is a key missing element to determine isospin-dependent kaon-nucleons scattering lengths, which have implications from particle and nuclear to astro-particle physics. The experimental effort takes place at INFN-LNF, Italy and at  J-PARC, in Japan, aiming at performing the measurement within the next few years.
The SIDDHARTA-2 experiment at the \Dafne{} collider of INFN-LNF has successfully completed its commissioning phase, SIDDHARTINO, in summer 2021. The final configuration is well underway  to start the 2022 kaonic deuterium run. To check the performance of the detector, the SIDDHARTA-2 collaboration collected 1.7 pb$^{-1}$ of kaonic helium data, which are presented in this article for the first time. Kaonic atoms transitions are visible in the SDD spectrum, showing state-of-the-art stable X-ray spectroscopic response and resilience to the hard radiation environment.

\section*{Acknowledgements}
The authors acknowledge C. Capoccia from INFN-LNF and H. Schneider, L. Stohwasser, and D. Pristauz Telsnigg from Stefan-Meyer-Institut für Subatomare Physik for their fundamental contribution in designing and building the SIDDHARTA-2 setup. We thank as well the
\Dafne{} staff for the excellent working conditions and permanent support.
\section*{Funding}
Part of this work was supported by the Austrian Science Fund (FWF): P24756-N20 and P33037-N; the Croatian Science Foundation under the project IP-2018-01-8570; EU STRONG-2020 project (grant agreement number 824093); the Polish Ministry of Science and Higher Education, grant number 7150/E-338/M/2018; and the Foundational Questions Institute and Fetzer Franklin Fund, a donor advised fund of Silicon Valley Community Foundation (grant number FQXi-RFP-CPW-2008).
This publication was made possible through the support of Grant 62099 from the John Templeton Foundation. The opinions expressed in this publication are those of the author(s) and do not necessarily reflect the views of the John Templeton Foundation.

\section*{References}
\bibliographystyle{unsrt}
\bibliography{siddharta2.bib}

\end{document}